\documentclass[a4paper,12pt]{article}
\usepackage{amsmath}
\usepackage{dsfont}
\usepackage{graphicx}

\begin{document}

\title{Spontaneous generation of chromomagnetic fields at finite temperature in the SU(3) gluodynamics on a lattice}
\author{Vadim Demchik\thanks{e-mail: vadimdi@yahoo.com}, Alexey Gulov\thanks{e-mail: alexey.gulov@gmail.com} and Natalia Kolomoyets\thanks{e-mail: rknv7@mail.ru}\\
{\small Dnipropetrovsk National University, 49010 Dnipropetrovsk, Ukraine}}
\maketitle

\begin{abstract}
The ~spontaneous generation ~of ~homogeneous chromomagnetic fields in the lattice SU(3) gluodynamics is investigated in the deconfinement phase of the model. A new approach based on direct measurements of the field strength on a lattice is developed. Vacuum magnetization is established by its influence on the probability density function of the simulated field strength. It is found that both the chromomagnetic fields corresponding to the diagonal SU(3) generators are simultaneously condensated and appear to be spatially co-directed. No vacuum magnetization is detected for the other SU(3) components. The temperature dependence of the spontaneously generated fields in physical units is fitted in the temperature interval 200 MeV -- 200 GeV as the usual power law with the anomalous dimension.
\end{abstract}

\section{Introduction}
The origin of cosmic magnetic fields in the early universe is one of the most actual problems of high energy physics. Numerous theoretical descriptions of large-scale magnetic field generation are discussed in the literature \cite{Grasso:2000wj,Elizalde:2012kz}. One of the elegant mechanisms is the spontaneous vacuum magnetization at high temperature. The possibility of the phenomenon has already been shown by both the analytical and numerical methods in the SU(2) gluodynamics \cite{Enqvist:1994rm,Skalozub:1996ax,Skalozub:1999bf,Demchik:2008zz}, in the SU(3) gluodynamics \cite{Skalozub:2002da}, in the standard model (SM) \cite{Demchik:2001zq}, and the minimal supersymmetric standard model (MSSM) \cite{Demchik:2002ks}. The stability of the spontaneously generated fields at high temperature was shown in Ref.~\cite{Skalozub:1999bf}.
Actually, the spontaneous vacuum magnetization is one of the distinguishable features of asymptotically free theories \cite{Skalozub:1999bf,Savvidy:1977as}.

In the SU(2) gluodynamics only one neutral magnetic field can be spontaneously generated, whereas in the SU(3) gluodynamics there are two possible spontaneously generated neutral chromomagnetic fields ($H_3$ and $H_8$ in case of the standard Gell-Mann matrices). The spontaneous vacuum magnetization phenomenon was investigated for the SU(2) gluodynamics on a lattice in \cite{Demchik:2008zz}, where the magnetic field was introduced with the twisted boundary conditions. The magnetized state was obtained by comparison of the free energies for the trivial vacuum state ($H_c=0$) and the magnetized vacuum state ($H_c\not=0$). In this paper we propose a new approach to study the effect. In contrast to setting external fields on a lattice, it should be reasonable to try to detect the presence of the condensed fields directly in the standard Monte Carlo (MC) techniques.

In \cite{Skalozub:2002da} the spontaneous generation of neutral SU(3) fields in the high-temperature limit $T\gg H^2$ was studied by analytical perturbative methods and the possibility of vacuum magnetization was shown. Nevertheless, the question about simultaneous generation of neutral chromomagnetic fields in the SU(3) gluodynamics still remains. In particular, the mutual spatial orientation of these fields should be qualified.

In the present paper the spontaneous vacuum magnetization at high temperature is investigated in the SU(3) gluodynamics on a lattice. We developed the procedure for detection of the condensate fields by direct analysis of lattice gauge configurations. This procedure is general for the deconfinement phase of any SU($N$) lattice gauge theory. It is grounded on the statistical analysis of chromomagnetic fields extracted from lattice configurations. The Cartesian components of quantum chromomagnetic fields have to be Maxwell distributed in simulations, whereas the presence of condensate fields should change this distribution. Relative spatial direction of the condensed chromomagnetic fields is a priori unknown and can be treated as a random quantity.
The complete electromagnetic tensor can be directly measured by taking the trace of a plaquette multiplied additionally by the gauge group generators. The deconfinement phase is spatially homogeneous, so the field tensor can be averaged over a lattice configuration to estimate the homogeneous vacuum state beyond quantum fluctuations. The statistical distribution of these averages is comparing with the Maxwell distribution to detect possible condensates. We show that the vacuum is magnetized for both the neutral fields $H_3$ and $H_8$ whereas for the other fields the vacuum magnetization is absent.

The paper is organized as follows. Sect.~2 provides the theoretical background of our approach to detect the condensed chromomagnetic fields. The results of Monte-Carlo simulations are described in Sect.~3. The temperature behavior of the condensate fields is derived and  compared with known results from the literature in Sect.~4. The discussion and possible applications are provided in Sect.~5.

\section{Detection of the chromomagnetic fields on a lattice}
We consider SU($N$) gauge theory on a lattice with the standard Wilson action
\begin{equation}\label{S}
 S_W=\dfrac{\beta}{N}\sum_{x\in\Lambda}\sum_{\mu<\nu}\mathrm{Re}\mathrm{Tr}(\mathds{1}-U_{\mu\nu}(x)),
\end{equation}
where $\beta={2N}/{g^2}$ is the inverse coupling, $\mathds{1}$ denotes the $N\times N$ identity matrix, and $U_{\mu\nu}(x)$ is the plaquette variable.
The sums run over all the lattice sites $x$ and directions $\mu,\nu$, and the isotropic lattice spacing $a$ is supposed.
The plaquette is constructed from the lattice field variables -- links,
\begin{equation}\label{u}
U_\mu(x)=\exp{(ia\lambda^bA^b_\mu(x))},
\end{equation}
where $\lambda^b$ is the $b$-th SU($N$) group generator, and $A^b_\mu(x)$ is the field potential in the continuous limit.

Links (\ref{u}) are subject to MC simulations. We implement standard boundary conditions: the sites on the opposite lattice edges are simply looped by the proper links.
The result of a MC run is a Boltzmann sequence of configurations with some numeric values for links. Then, any physical observable can be measured for each configuration and averaged over the Boltzmann ensemble.

We are interested to detect a kind of Savvidy vacuum in the deconfinement phase of the SU(3) gluodynamics. This means possible existence of a homogeneous field strength in the whole volume being a background of quantum fluctuations. In contrast to the confinement phase, which can be spatially fragmented to bound states, the deconfinement phase always occupy the whole lattice, so the homogeneous vacuum can be extracted as the field strength averaged over all the lattice sites. Actually, the huge lattices can show some domain structure, but relatively small lattices are expected to produce the condensate in some random direction.

The first step to detect a non-trivial vacuum state is the measurement of the field strength by a given lattice configuration. In fact, the field strength enters the plaquette variable used to construct the Wilson action (\ref{S}). Let us recall the plaquette definition:
\begin{equation}\label{umn}
 U_{\mu\nu}(x)=U_\mu(x)U_\nu(x+a\hat{\mu})U^\dag_\mu(x+a\hat{\nu})U^\dag_\nu(x),
\end{equation}
where $\hat{\mu}$ is the unit vector in direction $\mu$.
The expansion of link variables (\ref{u}) for small lattice spacing is well known (see for example \cite{Kogut:1979wt}):
\begin{equation}\label{pf}
 U_{\mu\nu}(x)=1+ia^2F^b_{\mu\nu}(x)\lambda^b-\frac12 a^4F^b_{\mu\nu}(x)F^{b'}_{\mu\nu}(x)\lambda^b\lambda^{b'} +\mathcal{O}(a^5).
\end{equation}
By applying the trace, one immediately obtains the field strength squared, which gives the correct continuous limit for the Wilson action.

As far as the Wilson action is considered as a good approach to field strength squared, we can use the same plaquette to extract the field strength (its flux through the plaquette, actually).
This could be done by multiplying plaquette (\ref{pf}) by generators $\lambda^b$, then taking the trace:
\begin{equation}\label{f}
a^2 F_{\mu\nu}^b(x) = -i\, \mathrm{Tr}\left[U_{\mu\nu}(x)\lambda^b\right] + \mathcal{O}(a^4).
\end{equation}
We will concentrate on the chromomagnetic part of the field tensor $F_{yz}=H_x$, $F_{xz}=-H_y$, $F_{xy}=H_z$.

The average of the chromomagnetic components of (\ref{f}) over a simulated lattice configuration gives us a direct measurement of the homogeneous chromomagnetic field generated spontaneously in the deconfinement phase, which will be called the {\it condensate} for brevity. In what follows we will mark these averaged values as $\vec{H}$,
where vector means spatial direction. These values are distributed within the Boltzmann ensemble with some probability density function $p(\vec{H})$  (p.d.f.). We suppose Cartesian components of $\vec{H}$ to be Gaussian with the variance $\sigma^2$, which is usual in MC simulations. The variance can be easily measured as the sample variance taken by the MC run. Then, the simulated field strength $\vec{H}$ is described by the following p.d.f.,
\begin{equation}\label{g}
 p(\vec{H})=\dfrac{1}{(2\pi)^{3/2}\sigma^3}\,e^{-(\vec{H}-\vec{H}_c)^2/2\sigma^2},
\end{equation}
where $\vec{H}_c$ is the condensate. The spatial direction of $\vec{H}_c$ appears to be unknown at each configuration generated in the MC run. It is random (at least, it has a random part), so p.d.f. (\ref{g}) could be written accurately as $p(\vec{H},\vec{H}_c)$. This is the reason against naive average of the Cartesian components of $\vec{H}$ over the run as a way to extract the condensate value from the generated Boltzmann ensemble.

Actually we are not interested in the random direction of $\vec{H}_c$, the question is what the absolute value of the condensate.
Using the spherical coordinates, we can rewrite probability in terms of the absolute values of the vectors, $H$ and $H_c$, as
\begin{equation}\label{gs}
 p(\vec{H})dH_x\,dH_y\,dH_z=\dfrac{1}{(2\pi)^{3/2}\sigma^3}\,\mathrm{e}^{-(H^2+H_c^2-2H_c H\cos\theta)/2\sigma^2}H^2dH\,d\phi\,d\cos\theta,
\end{equation}
where $\theta$ is the angle between $\vec{H}$ and $\vec{H}_c$, and $\phi$ is the azimuthal angle. Let us notice that the absolute value of the condensate $H_c$ is not random.

It is a well known fact, that the cosine of the angle between two random spatial vectors is distributed uniformly. It allows us to integrate
 (\ref{gs}) over angular variables obtaining the p.d.f. for the absolute value of the field strength:
\begin{equation}\label{d}
 p(H)=\dfrac{4\pi}{(2\pi)^{3/2}\sigma^3}\,H^2\mathrm{e}^{-H^2/2\sigma^2}\mathrm{e}^{-H_c^2/2\sigma^2}\dfrac{\sinh({HH_c}/{\sigma^2})}{HH_c/\sigma^2}.
\end{equation}
As a result, we get the value of the condensate $H_c$ as a parameter of the distribution of some directly measured variable. Thus, we can determine $H_c$ by analysis of the MC statistics obtained for $H$.

In case of condensate absence, $H_c=0$  and p.d.f. (\ref{d}) becomes the ordinary Maxwell distribution. A non-Maxwell shape of the p.d.f. means condensate generation. Since the p.d.f. contains only one unknown parameter $H_c$, it can be determined by fitting the theoretical expectation of the mean value by the corresponding sample mean from the MC run.

For qualitative analysis it is convenient to rewrite Eq.~(\ref{d}) in dimensionless (normalized) terms. Let us introduce the quantity
\begin{equation}\label{v0}
 H_0=\dfrac{4\sigma}{\sqrt{2\pi}},
\end{equation}
which is just the mean value of $H$ in case of $H_c=0$. Then, we can consider dimensionless variable $\eta$ with the following p.d.f.:
\begin{equation}\label{pHHc}
 \eta=\frac{H}{H_0},\qquad  p(\eta)=\frac{16 \eta}{\zeta\pi^{3/2}} \mathrm{e}^{-\zeta^2/4} \mathrm{e}^{-4\eta^2/\pi} \sinh\frac{2\zeta \eta}{\sqrt{\pi}},
\end{equation}
where $\zeta=H_c\sqrt{2}/\sigma$ depends on the condensate.

The mean value of $\eta$,
\begin{equation}\label{vt}
 \overline{\eta}=\frac{16}{\zeta\pi^{3/2}} \mathrm{e}^{-\zeta^2/4}\int\limits_0^\infty \eta^2\mathrm{e}^{-4\eta^2/\pi} \sinh\frac{2\zeta \eta}{\sqrt{\pi}}\,d\eta =f(\zeta),
\end{equation}
can be associated with its estimator from lattice simulations. To compute the condensate, $f(\zeta)$ must be inverted numerically. Function $f(\zeta)$ is plotted in Fig.~1.
\begin{figure}
\begin{center}
 \includegraphics[bb=0 1 258 170,width=0.75\textwidth]{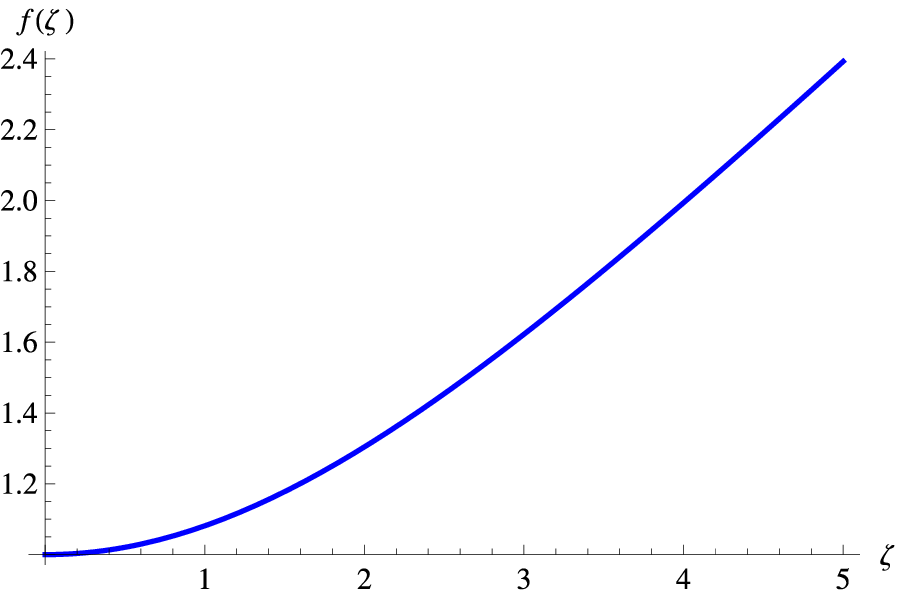}\\
\end{center}
  {\footnotesize {\bf{Figure 1:}}
Function $f(\zeta)$ from Eq.~(\ref{vt}) used to fit the p.d.f. of the field strength.}
\end{figure}
\begin{figure}
\begin{center}
  \includegraphics[bb=1 3 257 155,width=0.75\textwidth]{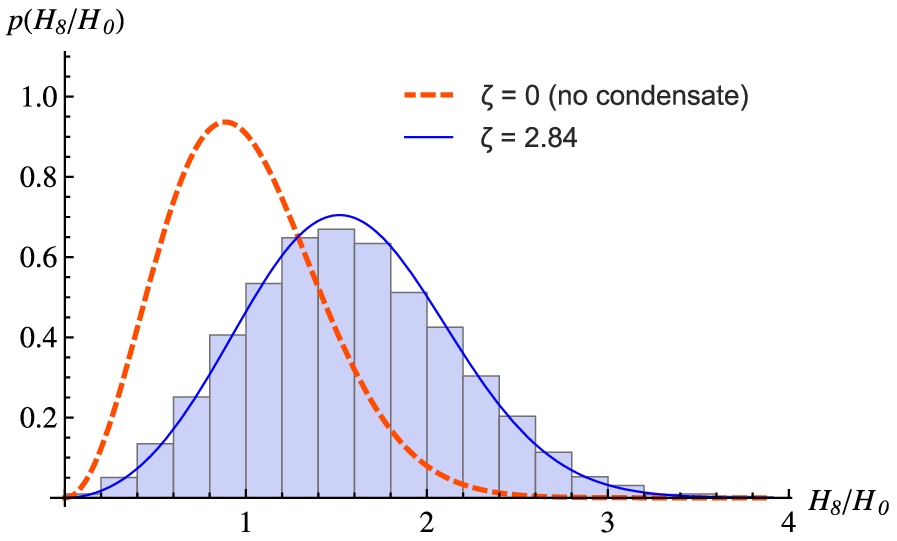}\\
\end{center}
  {\footnotesize {\bf{Figure 2:}}
The probability density function of the normalized field strength ($H_8/H_0$) obtained on a $18\times24^3$ lattice for $H_8$ field at $\beta=8$. The histogram corresponds to the lattice data (10\,000 measurements), the dashed line describes the absence of the condensate, and the solid line represents the p.d.f. fit assuming the condensate existence. The field $H_3$ behaves similarly to this plot.}
\end{figure}

In Figs.~2 and 3 we show an example of the described approach. The actual p.d.f. of the field strength is compared with the Maxwell distribution corresponding to the absence of the condensate ($H_c=0$ or $\zeta=0$, dashed lines). The parameter $\zeta$ is fitted by the sample mean obtaining either zero or non-zero value. We also check that the fitted value of $\zeta$ produces a good fit of the actual p.d.f. in case of condensate existence (solid lines correspond to (\ref{pHHc})). As it can be seen, the neutral SU(3) component $H_8$ has the condensate whereas there is no condensate for the charged component $H_1$. It should be noted that the condensation is also present for the neutral SU(3) field $H_3$ and absent for all the charged fields $H_2$, $H_4$, $H_5$, $H_6$ and $H_7$, as it can be expected from the theoretical reasons.
\begin{figure}
\begin{center}
  \includegraphics[bb=1 3 257 155,width=0.75\textwidth]{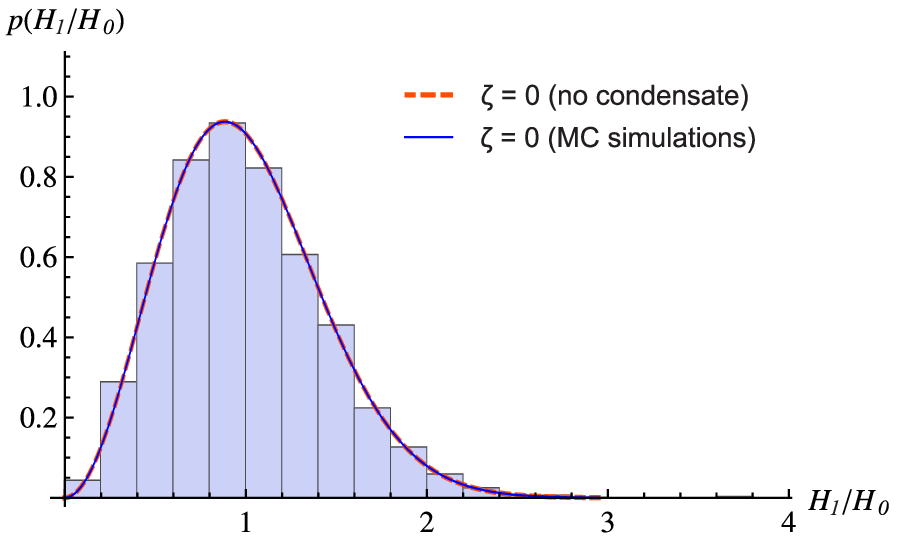}\\
\end{center}
  {\footnotesize {\bf{Figure 3:}}
The probability density function of the normalized field strength ($H_1/H_0$) obtained on a $18\times24^3$ lattice for $H_1$ field at $\beta=8$. The histogram corresponds to the lattice data (10\,000 measurements), the dashed line describes the theoretical expectation for the absence of the condensate, and the solid line represents the fitted p.d.f. which confirms the condensate absence. Fields $H_2$, $H_4$, $H_5$, $H_6$ and $H_7$ behave similarly -- they demonstrate the absence of condensates.}
\end{figure}

Thus, to find the spontaneously generated field strength one should measure the field tensor components $F_{xy}$, $F_{yz}$ and $F_{xz}$ in lattice simulations, average them over configurations ($H_x$, $H_y$, $H_z$), calculate their variance over the run ($\sigma^2$), compute the absolute values ($H=\sqrt{H_x^2+H_y^2+H_z^2}$) and their mean value over the run ($\overline{\eta}=\overline{H}/H_0$). Then, solving Eq.~(\ref{vt}) numerically with respect to $\zeta$, the condensate value $H_c=\sigma\zeta/\sqrt{2}$ can be derived.

It should be noted that the developed method is applicable for SU($N$) gauge group for any $N$ in the deconfinement phase.

\section{Monte-Carlo simulations}
The main aim of the paper is the investigation of chromomagnetic field condensates in the SU(3) gluodynamics. There are two neutral chromomagnetic fields in the model, $H_3$ and $H_8$, which correspond to the diagonal Gell-Mann matrices. Their condensates are expected to be generated at high temperatures.

The simulations are performed on the lattices of size up to $6\times32^3$ and $18\times24^3$ for various values of $\beta$ corresponding to the deconfinement regime. The lattice is updated by the heat-bath algorithm. To thermalize the system, $1000$ MC sweeps are used. All the measurements are performed on $10000$ configurations separated by $10$ sweeps. The MC kernel tries to update every lattice link on each sweep up to $10$ times.

MC simulations are performed on graphics processing units (GPU) on a HGPU cluster (AMD Radeon HD 7970, HD 6970, HD 5870 and NVidia GeForce GTX 560 Ti). The trivial parallelization scheme is implemented -- each GPU device simulates one lattice for a single parameter set per run. The subroutines for MC simulations on GPU (GPU kernels) are written in the Open Computing Language (OpenCL) \cite{OpenCL}. All the MC simulations and measurements are performed with the full double precision. All the necessary data are stored on GPU memory during a run, so the lattice size is restricted by the GPU device memory.

The ~Cartesian components ~of the chromomagnetic part of the field strength (\ref{f}) are measured at every working configuration for all the SU(3)-directions. Their averages over the lattice (denoted as $H_{a,i}$, $a=1,\ldots,8$; $i=x,y,z$) play a role of primary data for further statistical analysis.  The primary data are processed for every component separately in accordance with the approach described in Sect.~2. First, we compute the variance $\sigma_i$ over the run for each Cartesian component $H_i$. By averaging these variances over spatial directions we obtain the variance entering p.d.f. (\ref{g}), $\sigma = 1/3 \sum \sigma_i$. Then, the value of ${H}_0$ from Eq.~(\ref{v0}) can be found, which is used as a unit for $H$. Finally, we calculate the normalized absolute values $\eta=H/H_0=\sqrt{H_x^2+H_y^2+H_z^2}/H_0$ appeared in the run and analyze the distribution of the obtained sample.

To fit distribution (\ref{pHHc}) describing the condensate generation, we compute the average over the run (sample) $\overline{\eta}$, and find the parameter $\zeta$ as the numerical solution of Eq.~(\ref{vt}). After that, the histogram of the sample of $\eta$ is plotted to be compared with the fitted distribution (\ref{pHHc}) and the Maxwell distribution (the case of $\zeta=0$ meaning the condensate absence). An example of the histogram is shown in Fig.~1 for the lattice $18\times24^3$ and $\beta=8$ for the field $H_8$. The dashed line in Fig.~1 corresponds to the Maxwell distribution (no condensate), and the solid line represents the fitted distribution with the condensate. It can be seen that the histogram indicates the presence of the condensate rather then the condensate absence. Thus, the fitted parameter $\zeta$ gives the estimate of the field condensate,  $H_c=\sigma\zeta/\sqrt{2}$.

Analyzing all the SU(3) components of the chromomagnetic field tensor we find the condensates of neutral fields ($H_3$ and $H_8$), On the other hand, no condensates are detected for the rest of the SU(3) components.

In order to convert the obtained lattice values into the physical ones the standard relations are used,
\begin{equation}\label{lph}
 H=\dfrac{\varphi}{a^2},\qquad T=\frac{1}{aL_t},
\end{equation}
where $\varphi$ is the flux of the condensate field through the plaquette, $a$ is the lattice spacing and $L_t$ is the lattice size in the temporal direction. The values of the lattice spacing $a(g)$ have been computed using the standard technique described, for example, in \cite{Gattringer:2010zz}.

\begin{table}[ht]
\noindent
{\footnotesize {\bf{Table 1:}} The condensates of chromomagnetic fields for different lattices.}
\vskip 0.3cm
\begin{center}
\begin{tabular}{|c|c|c|c|c|c|c|}
\hline
\rule{0pt}{2.5ex}$\beta$&$a$,&$T$,&$\varphi_3/2$,&$\varphi_8/2$,&$H_{3c}$,&$H_{8c}$,\\
\rule{0pt}{2.5ex}&$\times10^{-2}$ fm&GeV&$\times10^{-3}$&$\times10^{-3}$&GeV$^2$&GeV$^2$\\
\hline
\multicolumn{7}{|c|}{\rule{0pt}{2.5ex}$12\times24^3$}\\
\hline
 8 &  1.994  & 0.825 & 3.716 & 2.239 &  0.7277 &  0.4385 \\
 9 & 0.6335  & 2.596 & 3.056 & 1.781 &   5.931 &   3.456 \\
10 & 0.2002  & 8.213 & 2.582 & 1.487 &   50.17 &   28.89 \\
11 & 0.06301 & 26.10 & 2.282 & 1.325 &   447.7 &   259.9 \\
12 & 0.01976 & 83.22 & 1.931 & 1.160 &   3853 &     2314 \\
\hline
\multicolumn{7}{|c|}{\rule{0pt}{2.5ex}$18\times24^3$}\\
\hline
 8 &   1.994 & 0.550 & 3.710 & 2.176 & 0.7266 & 0.4261 \\
 9 &  0.6335 & 1.730 & 3.108 & 1.797 &  6.032 &  3.486 \\
10 &  0.2002 & 5.476 & 2.649 & 1.589 &  51.47 &  30.88 \\
11 & 0.06301 & 17.40 & 2.241 & 1.299 &  439.7 &  254.9 \\
12 & 0.01976 & 55.48 & 1.987 & 1.144 &   3963 &   2283 \\
\hline
\multicolumn{7}{|c|}{\rule{0pt}{2.5ex}$6\times32^3$}\\
\hline
 8 &   1.994 & 1.649 & 3.686 & 2.239 & 0.7219 & 0.4384 \\
 9 &  0.6335 & 5.191 & 3.113 & 1.809 &  6.040 &  3.511 \\
10 &  0.2002 & 16.43 & 2.610 & 1.545 &  50.71 &  30.02 \\
11 & 0.06301 & 52.20 & 2.190 & 1.302 &  429.7 &  255.3 \\
12 & 0.01976 & 166.4 & 1.863 & 1.105 &   3716 &   2204 \\
\hline
\end{tabular}
\end{center}
\end{table}

The numerical results of Monte Carlo simulations for lattices $6\times 32^3$, $12\times 24^3$ and $18\times 24^3$ are presented in Table~1. In the first column of the table the inverse coupling $\beta$ is shown. The lattice spacing $a$ in physical units is given in the second column. The third column contains the temperatures $T$ in GeVs. The fluxes of condensate chromomagnetic fields through the plaquette are shown in the fourth and fifth columns. When all the quantities are measured in the energy units, the fluxes are dimensionless. The last two columns contain corresponding values of the condensate chromomagnetic fields $H_{3c}$ and $H_{8c}$ in GeV$^2$.

Another interesting question is the mutual spatial orientation of the condensate fields $\vec{H}_3$ and $\vec{H}_8$. The relative angle between the third and eighth fields can be found for every measurement in the run, since we know all the Cartesian components of the corresponding vectors. The cosine of the angle is
\begin{equation}\label{c38}
\cos\theta^*=\frac{H_{3,x}H_{8,x}+H_{3,y}H_{8,y}+H_{3,z}H_{8,z}}{H_3 H_8}.
\end{equation}
In Fig.~4 we show the histogram for $\cos\theta^*$ obtained for lattice $18\times 24^3$ at $\beta=8$. The fields seem to be co-directed, since the p.d.f. does not evidently fit the uniform distribution expected for two non-correlated vectors. To provide statistical estimates, let us note that random quantity (\ref{c38}) is expected to be distributed as the sample correlation for two Gaussian quantities. Namely, applying Fisher's transformation, we obtain Gaussian quantity $\mathrm{artanh}(\cos\theta^*)$, which mean and variance can be easily computed. In Fig.~4 we also present the fitted distribution, the mean value of $\mathrm{artanh}(\cos\theta^*)$ corresponds to $\cos\theta^*\simeq 0.829$ confirming the hypothesis of co-directed fields.

\begin{figure}\label{Fig:4}
\begin{center}
  \includegraphics[bb=0 1 258 162,width=4in]{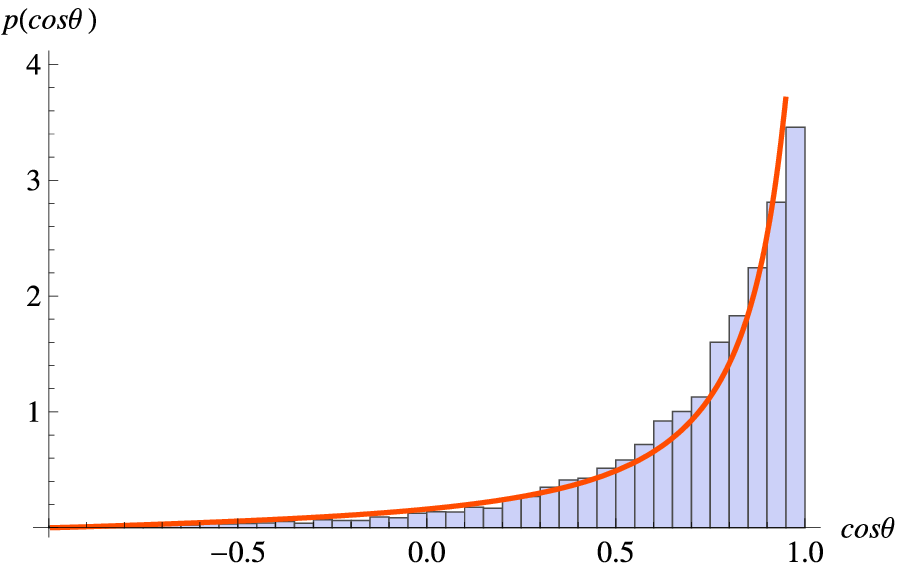}\\
\end{center}
{\footnotesize {\bf{Figure 4:}} Probability density function of the cosine of the angle between the condensates $\vec{H}_{3c}$ and $\vec{H}_{8c}$ on $18\times24^3$ lattice at $\beta=8$. The solid line represents the fit using Fisher's transformation.}
\end{figure}

\section{Temperature dependence of the field condensates}
In the previous section it was found that both the chromomagnetic fields ${H}_{3c}$ and ${H}_{8c}$ are generated in the SU(3) gluodynamics (see Table~1). It can be seen that the field condensate values are temperature dependent. In order to analyze this dependence, we plot the field strength versus the corresponding temperature squared. These plots for ${H}_{3c}$ and ${H}_{8c}$  for lattices $6\times 32^3$, $8\times 28^3$, $12\times 24^3$ and $18\times 24^3$ are shown in Fig.~5 in the logarithmic scale. The field strengths for various temperatures for each lattice size are plotted by different markers. The scatter plots demonstrate  linear dependence in logarithmic coordinates for various lattice sizes. The field condensates are practically independent of the lattice size in spatial and temporal directions for fixed $T$, so finite-size effects seem to play no crucial role in the observed effect.

\begin{figure}
\begin{center}
 \includegraphics[bb=0 3 254 246,width=0.43\textwidth]{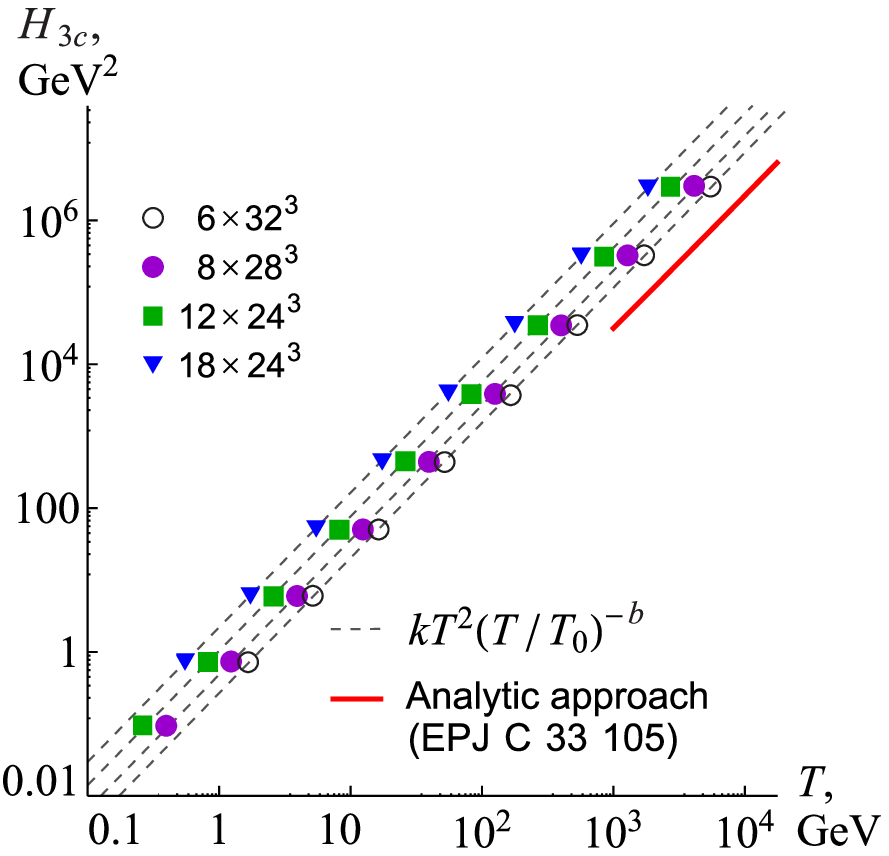}
 \hskip 0.075\textwidth
 \includegraphics[bb=0 3 254 246,width=0.43\textwidth]{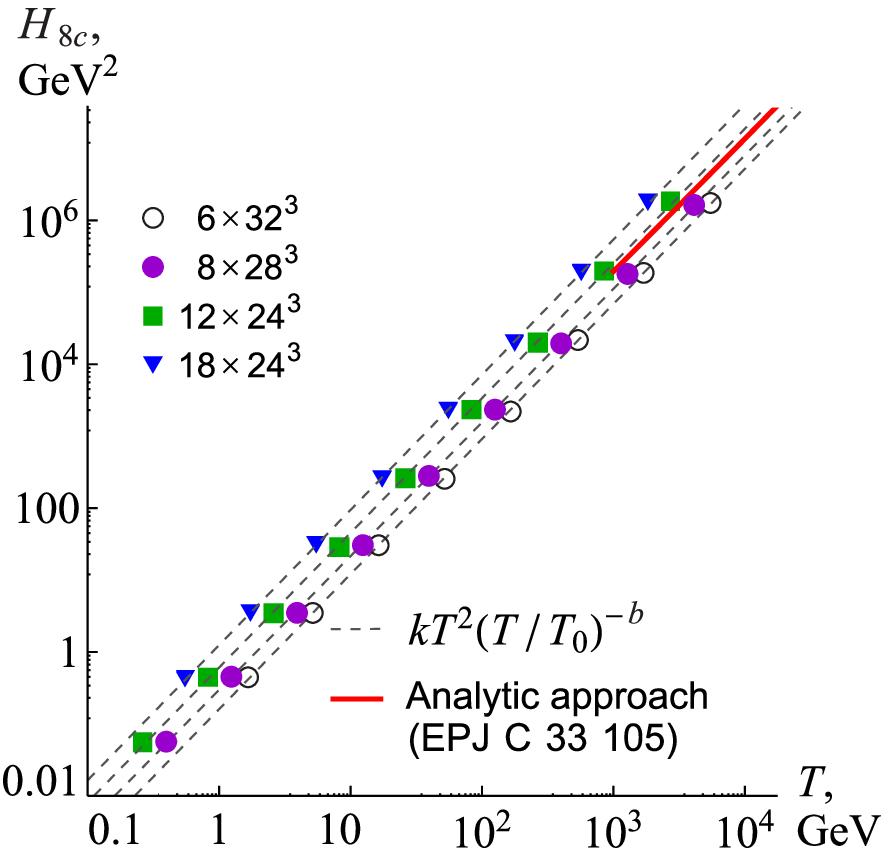}
\end{center}
{\footnotesize {\bf{Figure 5:}} $T^2$-dependence of the field strength ${H}_{3c}$ (left) and ${H}_{8c}$ (right). The dashed lines correspond to the data fits by Eq.~(\ref{pr}). The solid line corresponds to the analytic perturbative approach \cite{Skalozub:2002da}.}
\end{figure}

\begin{table}[ht]
\noindent
{\footnotesize {\bf{Table 2:}} Temperature dependence of the condensate fields fitted by the power law $H=kT^2(T/T_0)^{-b}$ with $T_0=0.2$\,GeV.}
\begin{center}
\nopagebreak
\vskip 0.3cm
\begin{tabular}{|c|c|c||c|c|}
\cline{2-5}
\multicolumn{1}{c|}{\rule{0pt}{2.5ex}} & \multicolumn{2}{c||}{$H_3$}&\multicolumn{2}{c|}{$H_8$}\\
\hline
lattice            & $k$  & $b$  & $k$  & $b$  \\
\hline
\rule{0pt}{2.5ex}  $6\times32^3$ & 0.33 & 0.12 & 0.20 & 0.12 \\
\rule{0pt}{2.5ex}  $8\times28^3$ & 0.60 & 0.13 & 0.38 & 0.14 \\
\rule{0pt}{2.5ex} $12\times24^3$ & 1.29 & 0.13 & 0.75 & 0.13 \\
\rule{0pt}{2.5ex} $18\times24^3$ & 2.65 & 0.13 & 1.52 & 0.12 \\
\hline
\end{tabular}
\end{center}
\end{table}

Obviously, the plotted temperature dependence of the field condensates can be fitted by the power law
\begin{equation}\label{pr}
 H_c=kT^2\left(\dfrac{T}{T_0}\right)^{-b},
\end{equation}
where $T_0$ is some arbitrary normalization scale. We choose $T_0=0.2$ GeV.

The fit results for each lattice size are shown in Table~2. The first column contains the information about lattices, the same as in Fig.~5. The fitted parameters $k$ and $b$ for the field ${H}_3$ are present in the second and third columns. The same parameters corresponding to the field ${H}_8$ are shown in the last two columns.
As it can be seen from Table~2, the slope parameters $b$ are very close for different lattices. This parameter is just `the anomalous dimension' of the field condensate, and it can be set to 0.13 for the SU(3) gluodynamics. The coefficient $k$, which defines the level of the fitting curve, grows with the increasing of temporal lattice size, but remains independent of the spatial one. This means, probably, either some finite-size effects, or systematic errors of the approach.

As it can be seen from above, both of the fields has similar magnitudes and identically grow in the investigated temperature range. No field dominates in the temperature range considered.

As it can be expected from the dimensional analysis, the dependence of the condensate field strength on the temperature is close to the quadratic function, ${H}\sim T^2$. The non-trivial behavior is described by the phenomenological law with the anomalous dimension $b$. It is interesting to compare the obtained results with analytic approaches assuming the continuous space-time. In \cite{Skalozub:2002da} the generation of chromomagnetic fields at high-temperature limit ($T\gg\sqrt{gH}$) was derived within the perturbative effective potential including the one-loop and daisy diagrams. Therein the following law was claimed
\begin{equation}\label{ss}
 {H}_{3c}=0.2976\dfrac{g^3T^2}{\pi^2},\hskip 2cm
 {H}_{8c}=1.8351\dfrac{g^3T^2}{\pi^2},
\end{equation}
where $g=g(T)$ is the running coupling. We can take the QCD running coupling to estimate the temperature behavior of $H$. At high temperatures the QCD coupling constant $\alpha=g^2/(4\pi)$ can be written as
\begin{equation}
 \alpha=\dfrac{1.38808}{\ln(T^2/(0.217\mbox{ GeV})^2)},
\end{equation}
where the numeric parameters were obtained by fitting the data from \cite{PDG}. The temperature dependence of ${H}_3$ and ${H}_8$ from (\ref{ss}) is shown in Fig.~5 by the solid lines. The slope and level of these lines are in a good agreement with the phenomenological law (\ref{pr}) from our MC results.

\section{Discussion}
In the paper spontaneous generation of chromomagnetic fields in the lattice SU(3) gluodynamics is investigated in the deconfinement phase. An approach for the detection of homogeneous condensate fields by direct measurements of the Cartesian components of the field tensor on a lattice is developed. The method is based on two steps. First, the homogeneous background field is found at each simulated configuration by averaging the components of the field tensor over the lattice. Second, the presence of condensate fields affect the statistical distribution of the extracted homogeneous background over the MC run. The p.d.f. of the absolute value of the background field depends only on the absolute value of the condensate field being distributed in accordance with the Maxwell distribution in case of no condensate fields. A non-Maxwell shape of the distribution indicates spontaneous generation of the condensate fields and allows to fit the value of the generated condensate. Probing different lattice geometries and various temperatures (see Table~1 and Fig.~5), we found the evident spontaneously generated magnetic components of the field tensor in the `neutral' SU(3) directions (both in the third and eight ones) whereas other SU(3) directions show the clear zero background. This serves as an additional argument that our approach reflects well motivated physics and cannot be interpreted as some random artificial lattice effects. In the physical units, the strength of condensate chromomagnetic fields monotonically grows with temperature increasing. We successfully fit the collected MC data by the phenomenological law with `anomalous dimension':
\begin{equation}\label{prdisc}
 H_{3c},H_{8c}\simeq T^2\left(\dfrac{T}{200\mbox{ MeV}}\right)^{-0.13}.
\end{equation}
This law incorporates the results obtained by analytic perturbative approach \cite{Skalozub:2002da} as a high-temperature asymptotics.

Both the neutral chromomagnetic fields $H_{3c}$ and $H_{8c}$ are simultaneously generated in the deconfinement phase in the whole temperature range investigated. These condensate fields have the same functional temperature dependence, the same order of magnitudes, so they evolve in a balanced manner. The direct analysis of the Cartesian components of the condensed fields shows that the fields are spontaneously generated in the same spatial direction.

The main results of the paper are obtained in physical units. It is checked that the condensate field strengths are independent of the spatial lattice size.

In general, the homogeneous chromomagnetic field in the whole space breaks the gauge invariance. Thus, the formation of domain structure at large scales is expected, which should restore the gauge invariance. So, the investigation of domain structure on huge lattices is also an interesting subject of further studies.

The method developed in Sect.~2 to determine the magnetized vacuum state is quite general. It can be applied for any SU($N$) lattice gauge theory. In this paper we restrict the investigation to the SU(3) gluodynamics as an important example.

In Sect.~4 we show that both the chromomagnetic fields $H_{3c}$ and $H_{8c}$ are simultaneously generated in the deconfinement phase in the wide temperature interval from 200 MeV to 200 GeV. These condensate fields have the same functional temperature dependence, so they evolve in a balanced manner. Moreover, both the fields appear to be spontaneously generated in the same spatial direction. It is derived in MC simulations that the spontaneous vacuum magnetization takes place only for $H_3$ and $H_8$ chromomagnetic fields ($H_{3c}\not = 0$, $H_{8c}\not = 0$), whereas there is no condensation of other chromomagnetic fields ($H_{1c},H_{2c},H_{4c},H_{5c},H_{6c},H_{7c}=0$).

In \cite{Demchik:2001zq} it was shown in the `one-loop plus daisy' perturbative approach that the chromomagnetic field $H_{3c}$ is generated in the standard model. The source of spontaneous magnetization effect is in the gauge sector of the model. The fermionic sector of the standard model slightly weakens the effect only.
The spontaneous vacuum magnetization was also studied in the MSSM \cite{Demchik:2002ks}, where the fermionic sector of the model is extended with the s-particles. It was obtain therein that doubled fermionic sector of the MSSM does not eliminate the effect.
As for our results, fermions should be accounted in further investigations.

\section*{Acknowledgements}
Authors are grateful to Vladimir Skalozub for useful discussions and suggestions.


\begin{thebibliography}{14}
\bibitem{Grasso:2000wj}
  D.~Grasso and H.~R.~Rubinstein,
  Phys.\ Rept.\  {\bf 348}, 163 (2001)
  [astro-ph/0009061].

\bibitem{Elizalde:2012kz}
  E.~Elizalde and V.~Skalozub,~
  Eur.\ Phys.\ J.\ C {\bf 72}, 1968 (2012)
  [arXiv:1202.3895 [hep-ph]].

\bibitem{Enqvist:1994rm}
  K.~Enqvist and P.~Olesen,
  Phys.\ Lett.\ B {\bf 329}, 195 (1994)
  [hep-ph/9402295].

\bibitem{Skalozub:1996ax}
  V.~V.~Skalozub,
  Int.\ J.\ Mod.\ Phys.\ A {\bf 11}, 5643 (1996).

\bibitem{Skalozub:1999bf}
  V.~Skalozub and M.~Bordag,
  Nucl.\ Phys.\ B {\bf 576}, 430 (2000)
  [hep-ph/9905302].

\bibitem{Demchik:2008zz}
  V.~I.~Demchik and V.~V.~Skalozub,
  Phys.\ Atom.\ Nucl.\  {\bf 71}, 180 (2008).

\bibitem{Skalozub:2002da}
  V.~V.~Skalozub and A.~V.~Strelchenko,
  Eur.\ Phys.\ J.\ C {\bf 33}, 105 (2004)
  [hep-ph/0208071].

\bibitem{Demchik:2001zq}
  V.~I.~Demchik and V.~V.~Skalozub,
  Eur.\ Phys.\ J.\ C {\bf 25}, 291 (2002)
  [hep-ph/0110280].

\bibitem{Demchik:2002ks}
  V.~I.~Demchik and V.~V.~Skalozub,
  Eur.\ Phys.\ J.\ C {\bf 27}, 601 (2003)
  [hep-ph/0208274].

\bibitem{Savvidy:1977as}
  G.~K.~Savvidy,
  Phys.\ Lett.\ B {\bf 71}, 133 (1977).

\bibitem{Kogut:1979wt}
  J.~B.~Kogut,
  Rev.\ Mod.\ Phys.\  {\bf 51}, 659 (1979).

\bibitem{OpenCL}
  A.~Munshi ed.,
  ``The OpenCL Specification'',
  Khronos OpenCL Working Group, (2011) 377 p.

\bibitem{Gattringer:2010zz}
  C.~Gattringer and C.~B.~Lang,
  Lect.\ Notes Phys.\  {\bf 788}, 1 (2010).

\bibitem{PDG}
  J.~Beringer {\it et al.}  [Particle Data Group Collaboration],
  Phys.\ Rev.\ D {\bf 86}, 010001 (2012).



\end{thebibliography}
\end{document}